\documentstyle[seceq,epsf,twocolumn]{jpsj}
\author{Akihisa {\sc Koga}, Seiya {\sc Kumada}
and Norio {\sc Kawakami}}
\title{
Quantum Phase Transitions in Two-Dimensional Spin Systems
with Ladder, Plaquette and Mixed-Spin Structures}
\inst{Department of Applied Physics,
Osaka University, Suita, Osaka 565-0871}

\recdate{February 4, 1999}

\abst{
Quantum phase transitions between the magnetically ordered
and disordered states are studied for the two-dimensional
antiferromagnetic quantum spin systems with ladder, plaquette,
and  mixed-spin structures. Starting with properly chosen
singlet-cluster configurations, we perform the series expansion
for the staggered magnetic susceptibility.  The phase boundary
is determined by applying the Dlog and biased Pad\'e approximants
to the staggered susceptibility thus obtained. The resulting phase
diagram allows us to discuss the quantum phase transitions
quantitatively, which agrees fairly well with the quantum Monte
Carlo results for several cases previously studied.
}

\kword{Heisenberg antiferromagnet, series expansion,
two dimension, quantum phase transition}

\def\runtitle{Quantum Phase Transitions in Two-Dimensional Spin Systems}

\def\runauthor{Akihisa {\sc Koga}, Seiya {\sc Kumada}
and Norio {\sc Kawakami}}

\begin{document}
\sloppy
\maketitle

\section{Introduction}

Recently low-dimensional spin systems with the gap for
the excitation spectrum have been the subjects
of considerable interest.
The spin gap is generated by various mechanisms, which may be
sensitive to the lattice structure, the competing interactions,
the topological nature of spins, etc.
A typical example is the spin plaquette system
such as $\rm Ca V_4 O_9$,
\cite{CaV4O91,meta1,CaV4O92,Troyer,KatoImadapla,GelCVO,CaV4O93,meta2}
which may be described by the two-dimensional (2D) Heisenberg model
on a 1/5 depleted square lattice or the meta-plaquette
model.\cite{meta1,meta2} The disordered ground state with the
spin gap observed in these systems may result from
the singlet-spin configuration formed in each plaquette.
Another prototypical example is the coupled spin-ladder
system\cite{Rice}
realized in the compounds such as SrCu$_{2}$O$_{3}$\cite{Azuma,k1}
and (Sr, Ca)$_{14}$Cu$_{24}$O$_{41}$\cite{Akimitsu}.
For such ladder systems with even number of legs, the spin gap is
generated to stabilize the disordered ground state. Interestingly
enough, the introduction of non-magnetic impurities into this system
induces the phase transition to the magnetically ordered
state.\cite{Azuma,k1,Fukuyama}
Furthermore, to the family of the spin gap systems we may add the
{\it mixed}-spin system, in which the topological nature of spins
as well as the lattice structure play an essential role to
produce the spin gap.  For instance, in one-dimensional
(1D) systems, a mixed-spin chain with the alternating array of
two kinds of spins has stimulated intensive experimental\cite{Yee}
and theoretical\cite{Pati,Vega,Tonegawa,Fukui,preprevious} studies.
It is known that the 1D mixed-spin chain realizes either the
ferrimagnetic or the singlet ground state depending on how we arrange
different type of spins on the 1D lattice.\cite{Pati,Vega,Tonegawa}

Various spin gap systems mentioned above provide us with an interesting
research area of quantum spin systems. Among others, the quantum phase
transition from the spin gap phase to the magnetically ordered
phase is one of the most interesting issues. Motivated by these hot
topics we investigated in the previous paper\cite{previous}
the competition between the magnetically ordered and disordered states
in the 2D quantum spin system which includes both of the bond- and
spin-alternations. By means of the non-linear sigma model and the
modified spin wave approach, we evaluated the spin gap and the
spontaneous staggered magnetization to discuss the quantum phase
transitions. Although this study enabled us to qualitatively describe
how the spin gap phase is driven to the antiferromagnetic one,
the resulting phase diagram was not sufficient enough to give
quantitative discussions.

The purpose of the present work is to study more quantitatively the
quantum phase transitions for the 2D spin systems with ladder,
plaquette, and mixed-spin structures. To this end we employ the
series expansion method developed\cite{dimer1,study} and extensively
used
by many groups for ladders,\cite{dimer2} 2D
systems,\cite{GelCVO,dimer3,Affleck}
Kondo lattice models,\cite{dimer4} bilayer systems,\cite{bilayer} etc.
In this approach, starting with properly
chosen singlet-spin clusters, we introduce the couplings among the
clusters perturbatively to carry out the series expansion of the
physical quantities such as the staggered susceptibility.
Then the critical point for the phase transition is determined by
applying the Pad\'e approximants\cite{Pade} to the physical quantities
thus obtained.

The paper is organized as follows. In \S 2, we first introduce
the Hamiltonian for 2D spin systems and outline how to apply
the series expansion techniques to our systems.
In \S 3, by performing the series expansion for the staggered
susceptibility  and then employing the Pad\'e approximants,
we obtain the phase diagram to discuss the competition
between the disordered and ordered states for three kinds of
2D quantum spin systems mentioned above. Brief summary is given
in the last section.

\section{Cluster Expansions}
We consider a 2D antiferromagnetic quantum spin system,
which is described by the following generalized Heisenberg Hamiltonian
\begin{eqnarray}
H & =  & H_0 + H_1, \\
H_0 & = & J_1 \sum_{(i,j)\in D_1}\mib S_{i} \cdot \mib S_{j}, \\
H_1 & = & J_2 \sum_{(i,j)\in D_2} \mib S_{i} \cdot \mib S_{j} +
J_3 \sum_{(i,j)\in D_3} \mib S_{i} \cdot \mib S_{j},
\end{eqnarray}
where $\mib S_j$ is the spin operator at the $j$-th site
on the square lattice and $J_{1}$, $J_{2}$, and $J_{3}$ are
the antiferromagnetic exchange couplings ($J_{1}, J_{2}, J_{3}>0$).
Note that the spin $\mib S_j$ is allowed to take different
values at each cite, which enables us to treat the systems
with mixed spins.

In order to apply cluster expansion techniques,\cite{study} the
Hamiltonian is divided into two parts. Since we start with a
strong-coupling singlet state, we take the unperturbed Hamiltonian
$H_{0}$ as an assembly of independent singlet-spin clusters formed by
the coupling $J_{1}$ (the corresponding bonds $(i,j)$ are sepecified by
$D_1$).
The interactions among independent clusters are
then taken into account by series expansions in the perturbed term
$H_{1}$.
We introduce two types of perturbations,
for which the corresponding sets of bonds $(i,j)$ are
denoted by $D_2$ and $D_3$.
The detail of $D_1, D_2$, and $D_3$ will be given
for each case in the following sections.
We henceforth assume that $\lambda(\equiv J_2/J_1) < 1$ and
$\alpha\lambda(\equiv J_3/J_1 )< 1 ( 0 < \alpha < 1 ) $.
We will see that this parameter regime indeed includes physically
interesting cases.

In the following,  we discuss the quantum phase transitions for the
2D antiferromagnetic spin systems with ladder, plaquette, and
mixed-spin structures. To this end, we introduce three kinds of
singlet-cluster configurations, i.e., the dimer singlet, the
plaquette singlet, and the mixed-spin-cluster singlet. The last one,
which is unique for our systems, is composed of a specific
singlet-spin configuration, {\it e.g.} $1/2 \circ 1 \circ 1/2$ (see
Figs. \ref{fig:model_a1} and \ref{fig:model_a2}).
The introduction of the mixed-spin-cluster allows us to perform a
systematic series expansion for 2D mixed spin systems.
Starting with  the above spin singlet states, we can carry out the
series expansion with respect to $\lambda$ as well as
$\alpha \lambda$. For the ladder, the plaquette, and the
mixed-spin structures, we refer to the corresponding expansions as
the dimer, the plaquette, and the mixed spin-cluster expansions,
respectively. We naturally expect that the introduction  of  $H_{1}$
perturbs the singlet ground state with the excitation gap and
gradually enhances the antiferromagnetic spin correlations,
finally giving rise to the long-range magnetic order.
This will be shown to be indeed the case for our systems.

We calculate the staggered spin susceptibility to determine the
critical point for the transition. We thus add the following
Zeeman term as a perturbation,
\begin{eqnarray}
H_{\rm ST} = h \left[\sum_{i \in A} S_i^{z} - \sum_{i \in B} S_i^{z}
\right],
\end{eqnarray}
where $h$ is the staggered magnetic field and $A$ ($B$)
denotes one of the two sublattices.
We estimate the ground state energy $E(h)$ of the total Hamiltonian
$H+H_{\rm ST}$ up to the second order in $h$, and then obtain
the magnetic susceptibility $\chi$ for the staggered field,
\begin{eqnarray}
\chi = - \left. \frac{ \partial^{2} E(h) }
{ \partial h^{2}}\right|_{h=0}.
\end{eqnarray}
This quantity is expanded as a power series in $\lambda$ as
\begin{eqnarray}
\chi = \sum_{n=0}a_{n}\lambda^{n}.
\end{eqnarray}
We calculate the susceptibility up to the eighth order in $\lambda$
for the dimer expansion and the fourth order for both of the plaquette
and the mixed-spin cluster expansions. We should recall here that
the phase boundary between the magnetically ordered and disordered
states is given by the critical line on which the staggered
susceptibility is divergent. Therefore a further approximate procedure
is necessary to deduce the sensible singularity by the asymptotic
analysis of the power-series expansion. For this purpose, we make use
of Pad\'e approximants\cite{Pade} for the susceptibility obtained up
to the finite order in $\lambda$. Besides ordinary Dlog Pad\'e
approximants, we also employ {\it biased} Pad\'e
approximants,\cite{Pade}
for which we assume that the phase transition in our 2D quantum spin
models
should belong to the universality class of the 3D
classical Heisenberg model.\cite{CHN} Then the critical value of
$\lambda_{c}$
is determined by the formula
$\chi \sim (\lambda-\lambda_{c})^{-\gamma}$
with the known exponent $\gamma=1.4$.\cite{Ferer} We shall
see below that the biased method provides a fairly good approximation
for $\lambda_{c}$ in some cases, and in general is useful to check
how well our Pad\'e approximants work.

\section{Quantum Phase Transitions}
In this section, we discuss the quantum phase transitions
by applying the series expansion to our models.
We have two parameters
in the perturbed term, $\lambda$ and $\alpha$,
so that we can observe
in different ways how the 2D antiferromagnetic correlations develop.
To confirm the validity of our series expansions and
Pad\'e approximants, we first study the 2D system with ladder structure
which was already studied extensively by various methods.
Our results are compared with those of the quantum Monte Carlo (QMC)
simulation in rather good agreement.\cite{KatoImada,ImadaIino,Nonomura}
We then move to the plaquette spin systems and the
mixed spin systems, and argue how the plaquette-singlet state
and the mixed-spin singlet state are driven to the
2D magnetically ordered state.

\begin{figure}[htb]
\vspace{0.1cm}
\epsfxsize=7cm
\centerline{\epsfbox{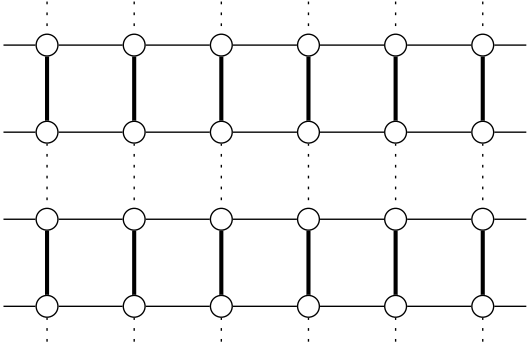}}
\caption{  2D spin system with the ladder structure.
  The circle represents the spin $s=1/2$. We refer to
the system as the coupled-ladder (coupled-dimer-chain) system,
when the bold, the thin, and the dashed lines represent the coupling
constants $1$, $\lambda_{\rm L} (\alpha_{\rm D}\lambda_{\rm D})$, and
$\alpha_{\rm L} \lambda_{\rm L} (\lambda_{\rm D})$,  respectively.}
\label{fig:model_d}
\end{figure}

\subsection{Ladder-structure systems}

Let us start with a 2D spin system with the ladder structure,
which is shown schematically in Fig. \ref{fig:model_d}.
In this figure, the circle represents
the $s=1/2$ spin sitting on the square lattice.
Note that this 2D model can be constructed from isolated dimers in two
ways depending on how we may introduce the couplings among dimers
perturbatively: we  may refer to the system correspondingly
as the coupled 2-leg ladders and the coupled dimer chains.
In the former case, the bold, the thin, and
the dashed lines in Fig. \ref{fig:model_d}
represent the coupling constants $1(=J_{1}), \lambda_{\rm L}(=J_{2})$,
and $\alpha_{\rm L}\lambda_{\rm L}(=J_{3})$, respectively.
By tuning the value of $\alpha_{\rm L}$, we naturally interpolate
the independent 2-leg ladders ($\alpha_{\rm L}=0$) and the 2D systems.
On the other hand, in the latter case,
by taking  $1(=J_{1}),  \lambda_{\rm D}(=J_{3})$,
and $\alpha_{\rm D}\lambda_{\rm D}(=J_{2})$, we can
investigate how the independent dimer chains ($\alpha_{\rm D}=0$)
are combined to make the 2D systems. We distinguish these two
constructions which may be complementary to each other
 because the available parameter
regime is restricted in our series expansion approach.

\begin{figure}[htb]
\vspace{0.1cm}
\epsfxsize=7cm
\centerline{\epsfbox{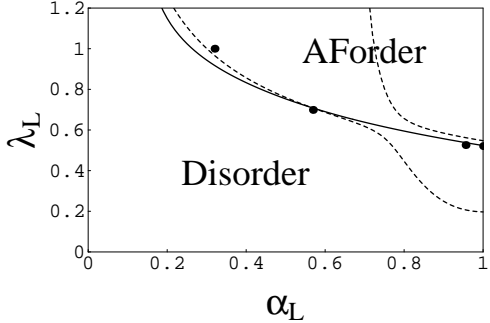}}
\caption{
Phase diagram for the coupled-ladder
system in Fig. \ref{fig:model_d}.
The solid (dashed) line indicates the phase boundary obtained by the
biased [3/4] (Dlog [3/4]) Pad\'e approximants.
The filled circles represent the results of the QMC simulations.
\cite{KatoImada,ImadaIino,Nonomura}}
\label{fig:res_d1}
\end{figure}

Let us first regard the 2D system as the coupled 2-leg ladders
as shown in Fig. \ref{fig:model_d}. The starting Hamiltonian
$H_{0}$ has the ground state with  the spin gap due to the
dimer singlets. We calculate the staggered susceptibility $\chi$
by means of the dimer expansion up to eighth order in $\lambda_{\rm L}$
for various values of $\alpha_{\rm L}$.
The resulting power series for some particular values of $\alpha_{\rm
L}$
are presented in Table I.
\begin{table}
\caption{Series coefficients $a_n$ for the dimer expansion of
the staggered susceptibility per site $\chi$
for the coupled-ladder system.}
\begin{tabular}{ccccc}
\hline\hline
n & $\alpha_{\rm L}=0.0$ & $\alpha_{\rm L}=0.2$ &
$\alpha_{\rm L}=0.5$ & $\alpha_{\rm L}=1.0$ \\
\hline
0 & 1.0000000 & 1.0000000 & 1.0000000 & 1.0000000 \\
1 & 2.0000000 & 2.2000000 & 2.5000000 & 3.0000000 \\
2 & 2.5000000 & 3.3350000 & 4.7187500 & 7.3750000 \\
3 & 1.7500000 & 3.7765000 & 7.7265625 & 17.062500 \\
4 &-0.44791667& 2.9802427 & 11.400106 & 37.401693 \\
5 &-2.7586806 & 1.3427618 & 16.016045 & 79.689670 \\
6 &-2.6446759 &0.32560333 & 22.218817 & 165.96349 \\
7 & 1.2087764 & 1.3087535 & 30.989600 & 340.66295 \\
8 & 5.9745629 & 3.6797174 & 43.252600 & 692.38191 \\
\hline\hline
\end{tabular}
\end{table}
Using Pad\'e approximants, we get
the phase diagram shown in Fig. \ref{fig:res_d1}.
In this figure, the solid (dashed) line represents the phase boundary
obtained by the biased [3/4] (Dlog [3/4]) Pad\'e approximants.
When  $\alpha_{\rm L}=0$, the system is reduced to the isolated
2-leg ladders with the interleg (intraleg) coupling
constant $1$ ($\lambda_{\rm L}$), which is known to have disordered
ground state with  spin gap.\cite{Rice}
Increasing the parameter $\alpha_{\rm L}$ with a fixed $\lambda_{\rm
L}$,
the antiferromagnetic correlation grows up,
and eventually the phase transition  to
the antiferromagnetically ordered state occurs.
For instance, if we determine the phase boundary
by means of biased [3/4] Pad\'e appoximants,
the critical value is given by $\alpha_{\rm L}= 0.26$ for $\lambda_{\rm
L}=1$.

We find that near $\alpha_{\rm L}\sim 0.8$ the phase boundary determined
by
the Dlog Pad\'e approximants exhibits a pathological behavior,
namely it branches out into upper and lower lines.
It may be obvious that these lines may not be physically sensible.
It is known that this type of pathology occasionally appears
in Pad\'e approximants.
If we discard these spurious parts in critical lines, we then find that
the results of
two Pad\'e approximants show the common behavior and are both in good
agreement
with those of the QMC simulations.\cite{KatoImada,ImadaIino,Nonomura}
Also, our results reproduce the critical value $\lambda_{\rm L}=0.54$
for $\alpha_{\rm L}=1$ which was previously obtained by Singh {\it et
al.}
\cite{dimer1}
We should notice, however, that the biased Pad\'e approximants may not
always give more accurate results than the ordinary Dlog Pad\'e
approximants. We must carefully determine the phase boundary after
trying various Pad\'e approximants, as will be momentarily shown below.

\begin{figure}[htb]
\vspace{0.1cm}
\epsfxsize=7cm
\centerline{\epsfbox{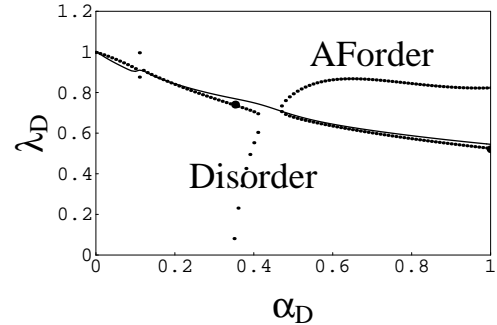}}
\caption{Phase diagram for the coupled-dimer-chain system
in Fig. \ref{fig:model_d}.  The solid (dotted) line indicates the
phase boundary obtained by the Dlog [4/3] (biased [4/3]) Pad\'e
approximants. The filled circles represent the QMC simulation results.
\cite{KatoImada,Nonomura}}
\label{fig:res_d2}
\end{figure}
We next regard the present system as the coupled dimer chains,
for which the corresponding couplings are defined
in Fig. \ref{fig:model_d}. Repeating a similar calculation as
in the previous case, we arrive at the phase diagram shown in
Fig. \ref{fig:res_d2}.
We list the obtained series for some values of $\alpha$ in Table II.
\begin{table}
\caption{Series coefficients $a_n$ for the dimer expansion of
the staggered susceptibility per site $\chi$
for the coupled-dimer-chain system.}
\begin{tabular}{ccccc}
\hline\hline
n & $\alpha_{\rm D}=0.0$ & $\alpha_{\rm D}=0.2$ &
$\alpha_{\rm D}=0.5$ & $\alpha_{\rm D}=1.0$ \\
\hline
0 & 1.0000000 & 1.0000000 & 1.0000000 & 1.0000000 \\
1 & 1.0000000 & 1.4000000 & 2.0000000 & 3.0000000 \\
2 &0.87500000 & 1.7750000 & 3.5000000 & 7.3750000 \\
3 &0.81250000 & 2.2865000 & 6.0312500 & 17.062500 \\
4 &0.77669271 & 2.9068927 & 10.016927 & 37.401693 \\
5 &0.74435764 & 3.6732624 & 16.341092 & 79.689670 \\
6 &0.71753608 & 4.6106279 & 26.245183 & 165.96349 \\
7 &0.69609122 & 5.7558135 & 41.690073 & 340.66295 \\
8 &0.67767823 & 7.1518929 & 65.617441 & 692.38191 \\
\hline\hline
\end{tabular}
\end{table}
In this figure, the solid and the dotted
lines represent the phase
boundaries obtained by the Dlog [4/3] and the biased [4/3] Pad\'e
approximants. In the latter analysis, we have two critical
lines, one of which (labeled by crosses) exhibits a pathological
behavior quite different from that of the Dlog Pad\'e
analysis.
If we discard this line, we find that
the results of two Pad\'e approximants show the
physically sensible behavior and are consistent with those of the
QMC simulations\cite{KatoImada,Nonomura}
(the filled circles in Fig. \ref{fig:res_d2}).

Finally we wish to note that at the point $(\alpha_{\rm D}, \lambda_{\rm
D})=(0,1)$,
our system just lies on the critical line which separates the ordered
and disordered states, as seen from Fig. \ref{fig:res_d2}. Since the
system in this case is reduced to the independent isotropic spin
chains, our numerical results reproduce the well-known
fact that the ground state of the spin-1/2 Heisenberg
chain is in a critical spin liquid phase, so-called Tomonaga-Luttinger
liquid phase, with neither the spin gap nor the long-range
order.\cite{Des}
We also note that the properties around this quantum critical point
have been already studied by Affleck {\it et al.}\cite{Affleck}

The above studies on the ladder-structure systems seem to give
rather satisfactory results, which encourage us to apply a similar
series expansion approach to the analysis of other quantum spin systems.

\subsection{Plaquette-structure systems}

\begin{figure}[htb]
\vspace{0.1cm}
\epsfxsize=7cm
\centerline{\epsfbox{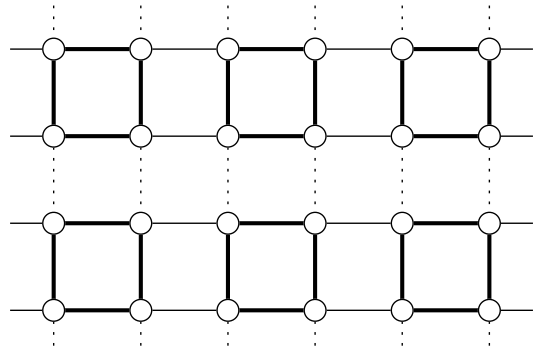}}
\caption{
2D spin system with the plaquette structure. The circle represents
the spin $s=1/2$. The bold, the thin, and the dashed lines represent
the coupling constants 1, $\lambda$, and $\alpha\lambda$.}
\label{fig:model_p}
\end{figure}

Let us now turn to the plaquette spin systems. We here consider two
kinds of the models. First, we are concerned with the system shown
in Fig. \ref{fig:model_p}.
The starting Hamiltonian $H_{0}$ describes a sum of the
independent plaquettes, whose ground state is spin singlet with
the excitation gap. Introducing the perturbed part $H_{1}$ may
induce the competition between the plaquette-singlet
correlation and the antiferromagnetic one.  By
this construction of the system, we naturally interpolate the
isolated plaquettes $\lambda=0$ (or isolated ladders $\alpha=0$) and
the normal square lattice.
\begin{figure}[htb]
\vspace{0.1cm}
\epsfxsize=7cm
\centerline{\epsfbox{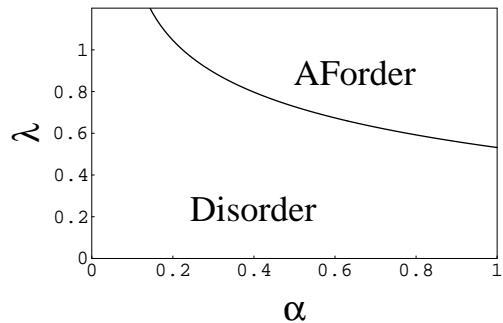}}
\caption{Phase diagram for the plaquette system
in Fig. \ref{fig:model_p}. The solid line indicates the phase
boundary obtained by the biased [1/2] Pad\'e approximants.}
\label{fig:phase_p}
\end{figure}
The coefficients are calculated up to the fourth order
in $\lambda$ by means of the plaquette expansion and
tabulated in Table III for some particular values of $\alpha$.
\begin{table}
\caption{Series coefficients $a_n$ for the plaquette expansion of
the staggered susceptibility per site $\chi$
for the 2D spin system with the plaquette structure.}
\begin{tabular}{ccccc}
\hline\hline
n & $\alpha=0.0$ & $\alpha=0.2$ &
$\alpha=0.5$ & $\alpha=1.0$ \\
\hline
0 & 1.3333333 & 1.3333333 & 1.3333333 & 1.3333333\\
1 & 1.7777778 & 2.1333333 & 2.6666667 & 3.5555556\\
2 & 1.6009195 & 2.6131044 & 4.3715197 & 7.9425797\\
3 & 1.1215579 & 2.9136374 & 6.8339620 & 17.102341\\
4 & 0.5614964 & 3.0347609 & 10.187867 & 35.146612\\
\hline\hline
\end{tabular}
\end{table}
Using the biased [1/2] Pad\'e approximants,
we obtain the phase diagram shown in Fig. \ref{fig:phase_p}.
In this phase diagram the system on the $\alpha =0$ axis
is composed of the assembly of the independent 2-leg ladders,
which belongs to the disordered spin-gap phase.
Away from this axis, the 2D antiferromagnetic correlation
grows up and the quantum phase transition to the ordered state occurs
at the critical value $\alpha_{c}$.
When $\lambda=1$, we find $\alpha_{c}=0.22$.
We note that this plaquette system with $\lambda=1$ is equivalent to
the coupled 2-leg ladders with $\lambda_{\rm L}=1$
discussed in the previous subsection,
where we have obtained the slightly different critical value
$\alpha_{\rm L}=0.26$.
It is also to be noticed that the mean-field theory by
the bond-operator representation and the QMC simulation
also yield the corresponding values $\alpha_c =0.25$\cite{bond-op} and
$0.32$\cite{ImadaIino} for the system with $\lambda=1$.
These results imply that
it may be necessary to carry out
higher order cluster expansions both for dimer and plaquette systems
to obtain more accurate critical values of $\alpha$
in the region $\lambda$
\raisebox{-0.3ex}{$\stackrel{\scriptstyle >}{\scriptstyle\sim}$} $1$
(This is indeed seen from Fig. \ref{fig:res_d1} for the ladder system).
Finally we wish to mention that for the special case ($\alpha=1$),
where each plaquette is connected with nearest neighbor
ones via the single coupling constant $\lambda$, similar results
have been reported by Fukumoto and Oguchi.\cite{Fukumoto}

\begin{figure}[htb]
\vspace{0.1cm}
\epsfxsize=7cm
\centerline{\epsfbox{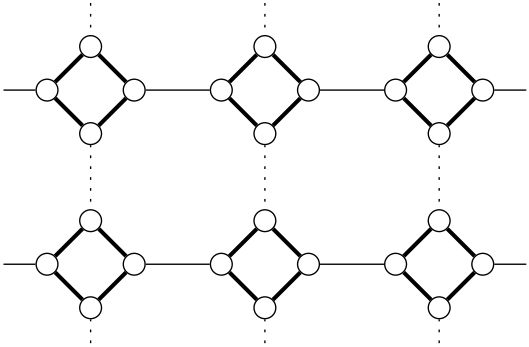}}
\caption{2D spin system composed of the plaquette chains.
The circle represents the spin $s=1/2$.
The bold, the thin, and the dashed lines indicate the coupling constants

1, $\lambda$, and $\alpha\lambda$.}
\label{fig:model_pc}
\end{figure}
\begin{figure}[htb]
\vspace{0.1cm}
\epsfxsize=7cm
\centerline{\epsfbox{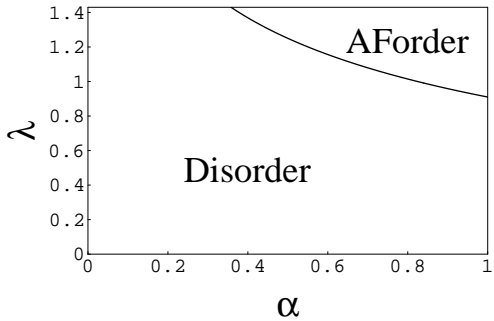}}
\caption{Phase diagram for the plaquette
system in Fig. \ref{fig:model_pc}.
The solid line represents the phase boundary obtained by
the biased [2/2] Pad\'e approximants.}
\label{fig:phase_pc}
\end{figure}

Let us next consider the system shown in Fig. \ref{fig:model_pc},
which may be considered to be made out of plaquette
chains,\cite{KatoImadapla,Ivanov,preprevious} because
this model, for $\alpha=0$ with finite $\lambda$,
is reduced to  the isolated chains with plaquette structures.
An interesting point is that this system is
topologically equivalent to the 1/5 depleted
square lattice \cite{CaV4O92,Troyer,KatoImadapla}
which is relevant to the study for the compound ${\rm Ca V_4 O_9}$.
The phase diagram shown in Fig. \ref{fig:phase_pc} is obtained by
the fourth order calculation and the biased [2/2] Pad\'e approximants.
The resulting series for some values of $\alpha$
are listed in Table IV.
\begin{table}
\caption{Series coefficients $a_n$ for the plaquette expansion of
the staggered susceptibility per site $\chi$
for the 2D spin system composed of the plaquette chains.}
\begin{tabular}{ccccc}
\hline\hline
n & $\alpha=0.0$ & $\alpha=0.2$ &
$\alpha=0.5$ & $\alpha=1.0$ \\
\hline
0 & 1.3333333 & 1.3333333 & 1.3333333 & 1.3333333\\
1 &0.88888889 & 1.0666667 & 1.3333333 & 1.7777778\\
2 &0.47985790 &0.73608925 & 1.1924150 & 2.1449010\\
3 &0.20889382 &0.47565756 & 1.0634199 & 2.6268926\\
4 &0.073161406&0.29653125 &0.92711496 & 3.1394401\\
\hline\hline
\end{tabular}
\end{table}
Let us first look at the results with the
value of $\lambda$ being fixed.  By increasing
the value of $\alpha $ from zero, we observe
the evolution of the plaquette chain to the
depleted square lattice.  For example, in the case of
$\lambda=1$, we get the critical value $\alpha_{c}=0.82$.
On the other hand, when we fix $\alpha=1$, we can see how
the isolated plaquettes are uniformly coupled to form
the 1/5 depleted square lattice, for which we obtain the critical
value $\lambda_c =0.91$. This value is in good agreement
with the result already obtained by QMC \cite{Troyer}
and also by the plaquette expansion.\cite{GelCVO}
As clear from these results, the plaquette structure
is quite essential, but is not sufficient to  produce the
spin gap for the isotropic 1/5 depleted square lattice
($\lambda =\alpha=1$), for which
we have the antiferromagnetic long-range order.
So, it is necessary to introduce
the dimer structure or frustrating couplings to have
the spin gap in this case.\cite{Troyer,GelCVO}

\subsection{Mixed-spin systems}
\begin{figure}[htb]
\vspace{0.1cm}
\epsfxsize=7cm
\centerline{\epsfbox{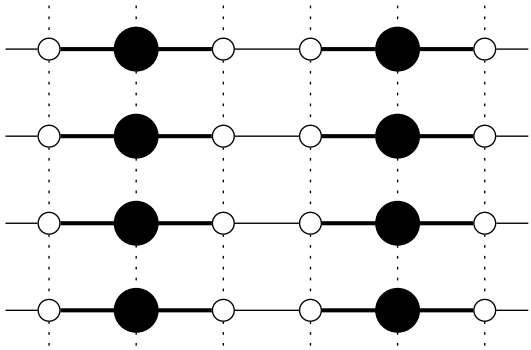}}
\caption{2D spin system with the {\it columnar} mixed-spin
structure.
The small (large filled) circle represents $s=1/2$ ($s=1$).
The bold, the thin, and the dashed lines indicate the coupling constants

$1$, $\lambda$, and $\alpha\lambda$.}
\label{fig:model_a1}
\end{figure}

\begin{figure}[htb]
\vspace{0.1cm}
\epsfxsize=7cm
\centerline{\epsfbox{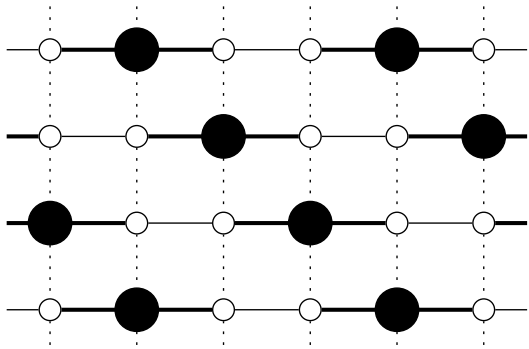}}
\caption{2D spin system with the {\it diagonal}
mixed-spin structure.
The small (large filled) circle represents $s=1/2$ ($s=1$).
The bold, the thin, and the dashed lines indicate the coupling constants

$1$, $\lambda$, and $\alpha\lambda$.
}
\label{fig:model_a2}
\end{figure}

So far, we have restricted our discussions to the 2D spin-1/2 systems,
for which the lattice structure plays an important role to generate
the spin gap. As mentioned in the introduction, the {\it mixed}-spin
systems with periodic array of different spins also have attracted
much attention both experimentally and theoretically.
In these systems, not only the lattice structure but also
the topological nature of spins become important for the gap formation.
In the previous papers,\cite{preprevious,previous} we have
investigated how the 1D mixed spin chains are coupled to
form the 2D mixed spin system
by means of the non-linear $\sigma$ model and the modified
spin wave analysis.  Although the quantum phase transitions between
magnetically  ordered and disordered phases were
described at least qualitatively by
the above approaches, the results obtained turned out to be
far from quantitative discussions.
The purpose in this subsection is to quantitatively explore
the phase transitions in 2D mixed spin systems
by using the series expansion method.
To be specific, we wish to deal with two typical systems
composed of $s=1/2$ and $s=1$, which are
shown in Figs. \ref{fig:model_a1} and \ref{fig:model_a2}.

\begin{figure}[htb]
\vspace{0.1cm}
\epsfxsize=7cm
\centerline{\epsfbox{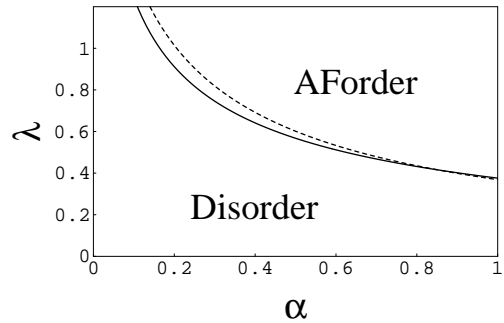}}
\caption{The phase diagram for the systems in Fig. \ref{fig:model_a1}
(the solid line) and Fig. \ref{fig:model_a2} (the dashed line).
The critical lines are determined by the biased [2/1]
Pad\'e approximants.
}
\label{fig:phase_mix}
\end{figure}

Let us first consider the system shown in Fig. \ref{fig:model_a1},
which we refer to as the "columnar mixed-spin system",
for which the mixed
spin chains are stacked uniformly in a vertical direction.
In this figure, the small circle (large filled circle)
represents $s=1/2 (s=1)$ and the bold, the thin, and the dashed lines
indicate the coupling constants $1(=J_{1}), \lambda(=J_{2})$,
and $\alpha\lambda(=J_{3})$, respectively.
When $\lambda=0 (H_{1}=0)$, the wave function of the ground state
is the direct product of mixed-spin-cluster
singlets given by the spin arrangement of $1/2 \circ 1 \circ 1/2$.
This is our starting configuration  for the cluster expansion.
The phase transition to the ordered state may be anticipated
when the values of $\alpha$ and $\lambda$ are increased.
We perform the mixed-spin cluster expansion up to the fourth order
and list the series coefficients for some particular values of $\alpha$
in Table V.
\begin{table}
\caption{Series coefficients $a_n$ for the mixed-spin cluster expansion
of
the staggered susceptibility per site $\chi$
for the {\it columnar} mixed-spin system.}
\begin{tabular}{ccccc}
\hline\hline
n & $\alpha=0.0$ & $\alpha=0.2$ &
$\alpha=0.5$ & $\alpha=1.0$ \\
\hline
0 & 1.7777778 & 1.7777778 & 1.7777778 & 1.7777778 \\
1 & 1.1851852 & 2.6074074 & 4.7407407 & 8.2962963 \\
2 &0.63981053 & 3.2769402 & 10.010760 & 28.642125 \\
3 &0.27852509 & 4.0253708 & 20.019578 & 88.590346 \\
4 &0.097548541& 4.7431208 & 38.497553 & 259.24615 \\
\hline\hline
\end{tabular}
\end{table}
Employing the biased [2/1] Pad\'e approximants for the
staggered susceptibility, we end up with the phase diagram shown
in Fig. \ref{fig:phase_mix} (solid line).
We find that the system at the point $(\alpha, \lambda)=(0,1)$, which
consists of independent mixed-spin chains, is in a disordered
phase with spin gap. This result correctly reproduces the fact that the
low-energy excitation in  the present mixed spin chain
has a gap, which is deduced via
the topological properties of the system.\cite{Fukui,preprevious}
We note that for $\lambda=1$ the phase transition to the ordered phase
occurs at the critical value $\alpha_c=0.16$.

Next we turn to the system shown in  Fig. \ref{fig:model_a2}.
We refer to it as the "diagonal mixed-spin system" since
the mixed  spin chains ($\alpha=0$)  are stacked diagonally.
The definition of the coupling constants is the same as that in
Fig. \ref{fig:model_a1}.
The mixed-spin cluster expansion up to the fourth order with
the biased [2/1] Pad\'e approximants yields
the phase diagram shown in Fig. \ref{fig:phase_mix} (the dashed line).
We tabulates the resulting series for some values of $\alpha$ in Table
VI.
\begin{table}
\caption{Series coefficients $a_n$ for the mixed-spin cluster expansion
of
the staggered susceptibility per site $\chi$
for the {\it diagonal} mixed-spin system.}
\begin{tabular}{ccccc}
\hline\hline
n & $\alpha=0.0$ & $\alpha=0.2$ &
$\alpha=0.5$ & $\alpha=1.0$ \\
\hline
0 & 1.7777778 & 1.7777778 & 1.7777778 & 1.7777778 \\
1 & 1.1851852 & 2.3703704 & 4.1481481 & 7.1111111 \\
2 &0.63981053 & 2.8229887 & 8.3587481 & 23.614326 \\
3 &0.27852509 & 3.1554307 & 15.787595 & 73.062503 \\
4 &0.097548541& 3.3658619 & 28.541914 & 216.50885 \\
\hline\hline
\end{tabular}
\end{table}
For $\alpha=0$, the system is correctly reduced to an assembly
of isolated mixed-spin chains which have the disordered ground state.
For $\lambda=1$ the phase transition to the ordered
phase occurs at the critical value $\alpha_{c}=0.21$.

Carefully observing Fig. \ref{fig:phase_mix} one notices that
two critical lines intersect each other in the vicinity of
$\alpha=0.87$.
In the region $\alpha < 0.87$,
the area of the disordered phase for the diagonal system
is larger than that for the columnar one.
This implies that when we increase $\lambda$ or $\alpha$
in this region, the antiferromagnetic correlation among spin clusters
grows up more easily in the columnar system than in the diagonal one,
because larger spins ($s=1$) are directly coupled
with each other in the columnar case, and stabilize the
long-range order more effectively.

On the other hand, when $\alpha>0.87$, the situation
is reversed, namely, the disordered phase in the columnar  system
is more stable than that for the diagonal system.
This may be understood by observing the behavior
in the limit $\alpha\rightarrow\infty$.
In this limit, the columnar system is reduced to the three-leg ladder,
\cite{preprevious} while the diagonal one still forms
the 2D network. Recall here that this three-leg
ladder consists of two $s=1/2$ chains and one $s=1$ chain,
which is known to have the spin gap.\cite{Fukui,preprevious}
Therefore, in the limit of $\alpha \rightarrow \infty$
the 2D antiferromagnetic correlation
vanishes for the  columnar system, while it may still survive for
the diagonal one. This may explain the behavior observed
in the region $\alpha>0.87$.

Up to now, the mixed-spin chains found experimentally are
known to exhibit the ferrimagnetic ground state.\cite{Yee}
It may be expected that the mixed-spin chains with singlet
ground state, as discussed here,
may be also synthesized experimentally in the future.
It would thus  be an interesting subject to observe
how such disordered systems are driven to the magnetically
ordered phase in the presence of the interchain couplings,
impurities, etc.

\section{Summary}
We have investigated the quantum phase transitions in the 2D spin
systems with ladder, plaquette, and mixed-spin structures.
In order to quantitatively study the phase transitions,
we have employed
the systematic cluster expansion methods.
It has turned out that the present approach improves to large extent
our previous results on the phase diagram obtained by
the non-linear $\sigma$ model as well as  the modified spin wave
analysis. For example, from the results on the ladder-structure
systems we have confirmed that the dimer expansion
analysis is comparable to the QMC simulation in some parameter regions.
We have also studied the phase diagrams for the spin systems with
the plaquette and the mixed-spin structures.
For plaquette systems, starting with isolated plaquettes
we have introduced the couplings among the plaquettes in two
distinct ways: the one is naturally extrapolated to the
2D square lattice, while the other is to the 1/5
depleted square lattice. In particular, for the latter case,
we have clarified how the plaquette
chains studied previously are coupled to form the 1/5 depleted
square lattice system. Also, for the mixed spin systems,
we have considered two ways for
stacking the spin chains, which are referred to as
the columnar and diagonal systems. It has been pointed out that
the stability of the disordered phase non-trivially
depends on how to stack the mixed spin chains.

In this paper we have concentrated on the staggered susceptibility
to establish the phase diagram.  Not only to confirm our
present results but also to get further information,
it is desirable to calculate the elementary excitation spectrum
in the same framework of the cluster expansion.
In this direction, we have performed a preliminary calculation
of the spin excitation gap for the columnar mixed spin system
up to the fifth order in the coupling constant.
The phase boundary determined in this way turns out to be
in fairly good agreement
with the one shown in Fig. \ref{fig:phase_mix}.
We thus believe that the present analysis of
the staggered susceptibility,
although it has been restricted to the fourth order, already
captures essential properties of the spin systems discussed in this
paper.

\section*{Acknowledgements}
We would like to thank A. Kl\"umper, H. Tsunetsugu and J. Zittartz
for useful discussions. The work is partly supported by a
Grant-in-Aid from the Ministry of Education, Science, Sports,
and Culture.

\end{document}